\documentclass[10pt, a4paper]{article}

\usepackage{indentfirst}
\usepackage{misccorr}
\usepackage{graphicx}
\usepackage{amsmath}
\usepackage{pstool}
\usepackage{hyperref}
\usepackage{amssymb}
\usepackage{tocbibind}
\usepackage[normalem]{ulem}
\usepackage{authblk}

\makeatletter
\renewcommand{\appendix}
{
\par
\setcounter{section}{0}

}

\makeatother

\hypersetup{unicode=true}

\begin{document}

\title{Self-Excited Ising Game \footnote{Corresponding author: A.~Antonov}
}
\author{A.~Antonov$^{(1)}$\footnote{antonov@lpi.ru \\ \\ \\ Preprint submitted to Physica A} }
\author{A.~Leonidov$^{(1,2)}$}
\author{A.~Semenov$^{(1,2,3)}$}
\affil{{\small (1) P.N. Lebedev Physical Institute, Moscow, Russia\\
		(2) Moscow Institute of Physics and Technology, Dolgoprudny, Russia\\
		(3) Higher School of Economics, Moscow, Russia}}
\date{}

\maketitle

\begin{abstract}
	Effects of dynamical activity spillover in a noisy binary choice game (Ising game) on a complete graph are studied. Binary choice games are very important for both economics and statistical physics playing a role of the bridge between these two fields. In this paper we investigate the effects of self-excited activity induced by activity spillover on relaxation to equilibria and transitions between metastable equilibria at finite times. Using the formalism of master equations we show that both relaxation and interequilibria transitions at finite time are accelerated by the effects of activity spillover.
    
\end{abstract}


	\maketitle

\newpage

\section{Introduction}

Studies of noisy binary choice games are of special interest because of the existence of close parallels to statistical physics of spin systems, in particular to static and dynamic properties of phase transitions in them \cite{Blume2003,Bouchaud2013,salinas2001introduction}. These parallels are particularly intriguing because of the fundamentally different  origins of equilibria in game theory and statistical physics: in game theory equilibration is a result of balancing individual interests while in statistical physics equilibration is a search of a global minimum of free energy. For the noisy binary choice problem on complete graphs it is long known, see \cite{Blume2003} and references therein, that for a special choice of noise (that of the Gumbel distribution) static game-theoretic equilibria are defined by the  mean-field Curie-Weiss equation, see e.g. \cite{salinas2001introduction}, describing phase transitions in magnetics. Recently in  \cite{leonidov2019quantal} it was established that static game-theoretic equilibria in noisy binary choice games on graphs correspond to the so-called quantal response equilibria in game theory \cite{goeree2016quantal}.

The dynamics of games can, however, be fundamentally different from conventional spin dynamics due to a variety of possible mechanisms. One of these is a possibility of self-excitation that was intensively studied for so-called Hawkes processes \cite{Hawkes1971}, in which agent's activity is boosted by her activity in the past. The effects of Hawkes self-excitation were considered in applications to finance \cite{sornette,article12}, characteristics of earthquakes \cite{Ogata1988} and other subjects, see the recent review in \cite{Laub2015}. Recently a master equation formalism for Hawkes processes was developed in \cite{kanazawa2020field,kanazawa2020non}. In the papers \cite{chevallier} and \cite{hainaut} more complicated versions of such equations were considered.

The main idea of the present paper is studying the effects of an activity spillover fundamentally different from the Hawkes self-excitation mechanism having its origin in the amplification of an agent's activity through activity spillover induced by the past activity {\it of other agents} for a noisy binary choice game (Ising game) on a complete graph. 

The outline of the paper is as follows. In the Section~\ref{isg} we start with describing a best-response type dynamics of a noisy self - excited Ising binary choice game. We then proceed with deriving a master equation describing the dynamical evolution of the system and derive evolution equations for average choice and choice intensity generalising the well-known equations fro Glauber dynamics in the Ising model on a complete graph.  It shown that self-excitation leads to acceleration of convergence to the stable asymptotic equilibria of the game. In the Section \ref{trans} we consider the effects of activity spillover on the rate of transitions between (meta)stable equilibria at finite times. We show that self-excitation leads to an exponential amplification of the rate of such transitions. In the Section~\ref{concl} we list our conclusions.

\section{Self-excited dynamical Ising game}\label{isg}

\subsection{Game description}

In what follows we consider a dynamical noisy binary choice game of $N$ agents on a graph  $\mathcal G$. The strategy space of each agent $i$ includes two strategies $s_i=\pm1$ so at given time $t$ the system is fully characterised by the set ${\bf s}_t = (s_1, \ldots , s_n)_t$. The temporal evolution of the strategies configuration ${\bf s}_t \to {\bf s}_{t+\delta t}$ within a small time interval $dt $ is assumed to be driven by a strategy flip  $s_i \to -s_i$ of some agent $i$:
\begin{equation}
(s_1, \ldots , s_i, \ldots, s_n)_t \to (s_1, \ldots , -s_i, \ldots, s_n)_t
\end{equation}
The flip probability $\pi_{i,t}$ within a time interval $(t,t+\delta t)$ is assumed to have a form
\begin{equation}
\pi_{i,t} = \lambda_i (t)  \delta t \; \gamma (s_i \to -s_i \vert {\bf s}_{-i,t})
\end{equation} 
where $\lambda_i(t) \delta t$ is a time-dependent probability of having a possibility for an agent $i$ to reconsider a strategy within a time interval  $(t,t+\delta t)$ and $ \gamma (s_i \to -s_i \vert {\bf s}_{-i,t})$ is a probability of changing a strategy dependent of the current configuration ${\bf s}_{-i,t}$ of strategies in the neighbourhood $\mathcal{V}_i$ of the node $i$. Let us denote the sequence of flip times of an agent $i$ by  $\{t_i^{k}\}_{i \in \mathbb {N}}$. The main effect studied in the present paper is the acceleration of the flip activity of the node $i$ by the preceding flip activity of its neighbours so that
 \begin{equation}\label{eq1}
    \displaystyle \lambda_i(t)=\lambda_0 +\sum_{k \in \mathcal{V}_i}\sum_{t_j^k<t}h(t-t_j^{(k)}),
\end{equation}
where $h(t-t')$ is a memory kernel. Equation \eqref{eq1} describes flip activity spillover from the neighbours of a node and is therefore a generalisation of the Hawkes activity spillover which is self-induced. The case of constant intensity 
$\lambda_i(t)=\lambda_0 $ corresponds to a standard poissonian dynamics underlying the Ising game, see e.g. \cite{Blume2003,leonidov2019quantal}.

In what follows we shall assume a noisy best response (Ising-Glauber) flip rate\footnote{This choice corresponds to the Gumbel noise in the individual agents utilities.}
\begin{equation}
   \displaystyle \gamma_i(t)=\frac{1}{2}\left[1-s_i\tanh\left(\beta \mathcal{J}\sum_{k \in \mathcal{V}_i} s_k(t) \right)\right],
   \label{defgamma1}
\end{equation}
where $\beta=1/T$ is an inverse temperature and $\mathcal{J}=J/N$ is an Ising coupling constant\footnote{The $1/N$ factor is choisen in analogy with the corresponding choice in the Ising model on a complete graph in which it ensures an appropriate free energy scaling. In game theory this choice is optional.} and a markovian exponential memory kernel
\begin{equation}\label{eq2}
h(t)=\mu e^{-b t}
\end{equation}

In the present study we consider a complete graph topology in which at large $N$ equation \eqref{defgamma1} takes the form \cite{Bouchaud2013, Blume2003}:
\begin{equation}
  \displaystyle \gamma (t)=\frac{1}{2}\left[1-s_i\tanh\left(\beta J m (t) \right)\right] \;\; \rightarrow \;\; \gamma_{\pm} (m)=\frac{1}{2}\left(1 \pm \tanh(\beta J m)\right)
   \label{defgamma2}
\end{equation}
where $\gamma_{\pm} = \gamma(\mp s \to \pm s)$ and 
\begin{equation}
m(t) = \frac{1}{N} \sum_{i=1}^N s_i 
\end{equation}

As for the memory kernel, it turns out convenient to change normalisation so that
\begin{equation}\label{eq4}
    \lambda(t)=\lambda_0 +\frac{\mu}{N} \sum_{\tau_k}e^{-b(t-\tau_k)}
\end{equation}
As all vertices in the complete graph are equivalent in describing the system evolution it is natural to use a collective intensity of the flip process  $\Lambda (t) = N \lambda (t)$.

\subsection{Dynamical evolution}

A state of the system at any given time is fully described by the values $(\Lambda (t),m(t))$. Their evolution in the time interval  $[t; t+\delta t)$ is driven by the two following mechanisms:
\begin{itemize}
\item The strategy flips $-1 \to 1$ or $1 \to -1$ leading to a constant shift in the process intensity $\Lambda \to \Lambda + \mu$ and a change in the average strategy $m \to m \pm 2/N$ taking place with the following probabilities $\pi_{\pm}$
\begin{equation}
\left(\Lambda, m\right) \to \left(\Lambda+\mu, m \pm \frac{2}{N}\right) \;\; \leftrightarrow \;\; \pi_{\pm} = \Lambda \delta t\cdot\gamma_{\pm}(m)\frac{1 \mp m}{2}
\label{pipm}
\end{equation}
\item Decrease of the rate $\Lambda$ taking place in the absence of a strategy flip
\begin{equation}
\left(\Lambda, m\right) \to \left((\Lambda-\Lambda_0)e^{b\delta t}+\Lambda_0, m\right)
\label{pi01}
\end{equation}
occurring with the probability
\begin{equation}
\pi_0 = 1-\Lambda \delta t\left(\gamma_{-}(m)\frac{1+m}{2}+\gamma_{+}(m)\frac{1-m}{2} \right)
\label{pi02}
\end{equation}
\end{itemize}

From equations (\ref{pipm},\ref{pi01},\ref{pi02}) we obtain the following  master equation for the probability density function $P\left(\Lambda, m; t\right)$

\begin{eqnarray}
\frac{\partial P(\Lambda,m;t)}{\partial t} & = & b\frac{\partial}{\partial \Lambda}\big((\Lambda-\Lambda_0 ) P(\Lambda,m;t)\big) \nonumber \\
& + &\Bigg(\left(\Lambda-\mu\right)\left(\frac{m+1}{2}+\frac{1}{N}\right)\gamma_{-}\left(m+\frac{2}{N}\right)P(\Lambda-\mu,m+\frac{2}{N};t) \nonumber \\
& + & \left(\Lambda-\mu\right)\left(\frac{1-m}{2}+\frac{1}{N}\right)\gamma_{+}\left(m-\frac{2}{N}\right)P(\Lambda-\mu,m-\frac{2}{N};t) \nonumber \\
& - & \Lambda\left(\frac{m+1}{2}\gamma_{-}(m)+\frac{1-m}{2}\gamma_{+}(m) \right)P(\Lambda, m; t)\Bigg)
\label{eq7}
\end{eqnarray}
Let us note that if one neglects evolution of $m$, equation \eqref{eq7} coincides with the master equation for Poissonian self-excited processes obtained in \cite{kanazawa2020field}.

From \eqref{eq7} a standard computation in the limit $N \to \infty$ leads us  to the following evolution equations for $m$ and $\lambda$:

\begin{eqnarray}
\dot{m}(t) & = & -\lambda (t) \left[ m(t)-\tanh(\beta J m(t)) \right]  \label{evm} \\
\dot{\lambda}(t) & = & \lambda(t) \left[ 1-m(t)\tanh(\beta J m(t)) \right]  -b \left[ \lambda(t)-\lambda_0 \right], \label{evel}
\end{eqnarray}
where we have performed the following rescaling of the variables: $$\lambda \to \frac{2\lambda}{\mu}, \lambda_0 \to \frac{2\lambda_0}{\mu}, b\to \frac{2b}{\mu}, t \to \frac{\mu t}{2}$$.

Equation \eqref{evm} is nothing else but the standard evolution of the average choice on a complete graph \cite{Blume2003,Bouchaud2013,leonidov2019quantal} that is also equivalent to the mean-field evolution equation for the Ising model \cite{salinas2001introduction} with the time-dependent intensity $\lambda(t)$. From equation \eqref{evm} it is clear that the stationary configurations of the Ising game under consideration are the same as in the constant intensity case $\lambda = \lambda_0$ and are described by the Curie-Weiss equation \cite{Blume2003,Bouchaud2013,leonidov2019quantal} 
\begin{equation} 
m_{\rm eq} = \tanh(\beta J m_{\rm eq}),
\label{CW}
\end{equation}
the difference with the standard case being due to the temporal evolution of relaxation intensity described by equation \eqref{evel}. The equilibrium configurations $m_{\rm eq}(\beta)$ described by the corresponding solutions of \eqref{CW} are therefore the conventional ones: $m_{\rm eq} = 0$ at high temperatures $\beta J <1$ and $m_{\rm eq} = \pm m_0 (\beta)$ at low temperatures $\beta J>1$.

As to the characteristic regimes of evolution of the process intensity, it is convenient to consider the vicinity of equilibrium $m \sim m_{eq}$ in which from equations \eqref{evel}, \eqref{CW} there follows that
\begin{equation}
\dot{\lambda}(t) \simeq \left[ 1 - m_{\rm eq}^2 -b\right] \lambda(t)+b\lambda_0 
\end{equation}
so that
\begin{equation}
\lambda(t) = \frac{\lambda_0}{1-m_{\rm eq}^2 -b}  \left[ (1-m_{\rm eq}^2 ) e^{(1-m_{\rm eq}^2 -b)t} -b\right]
\end{equation}
The two characteristic regimes corresponding to growing and relaxing intensity are thus characterised by the following asymptotic behavior:
\begin{eqnarray}
1-m_{\rm eq}^2 -b >0  & \to & \left. \lambda (t) \right \vert_{t \to \infty} \sim  \frac{e^{(1-m_{\rm eq}^2 -b) t}}{1-m_{\rm eq}^2 -b} \lambda_0 \nonumber \\
 1-m_{\rm eq}^2 -b <0   & \to & \left. \lambda (t) \right \vert_{t \to \infty} \; \sim \frac{b}{m_{\rm eq}^2+b-1} \lambda_0
\end{eqnarray}

From the derived system of evolution equations \eqref{evm},  \eqref{evel} it is clear that the main effect of time-dependent intensity should be in speeding up relaxation to the appropriate temperature - dependent equilibrium. A nontrivial part of this effect is its dependence on the characteristic memory timescale $\tau_\lambda = 1/b$. At parametrically small $\tau_\lambda$ (large $b$) one expects a rapid recovery of the base intensity $\lambda_0$ while at large $\tau_\lambda$ (small $b$) one, on the contrary, expects a prolonged period of high-intensity evolution. Therefore at large $b$ the influence of self-excitement in the game development should be small while at large  $b$ it should, on the contrary, be large.

\begin{figure}[htp!]
\begin{minipage}[h]{0.47\linewidth}
\center{\includegraphics[width=1\linewidth]{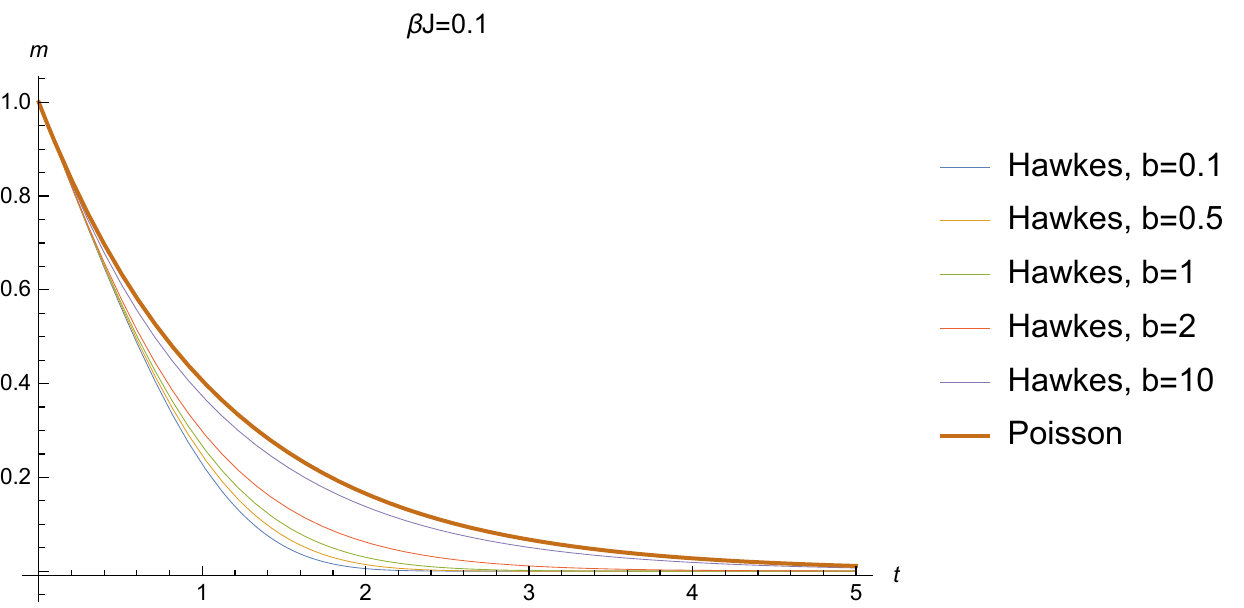}} a) \\
\end{minipage}
\hfill
\begin{minipage}[h]{0.47\linewidth}
\center{\includegraphics[width=1\linewidth]{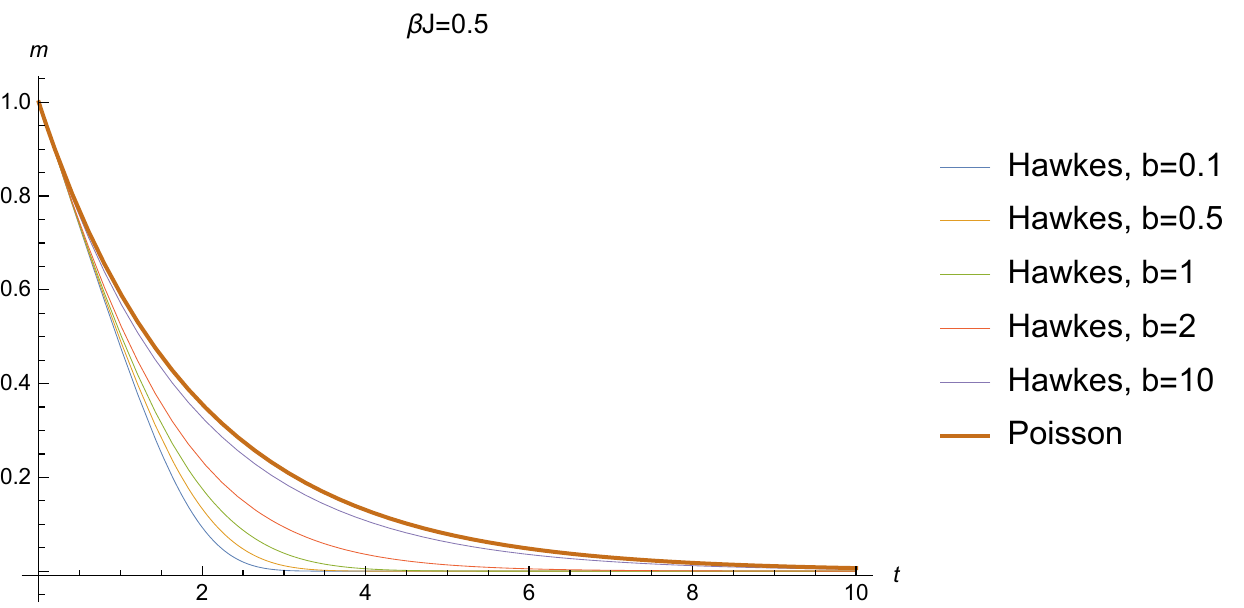}} \\b)
\end{minipage}
\vfill
\begin{minipage}[h]{0.47\linewidth}
\center{\includegraphics[width=1\linewidth]{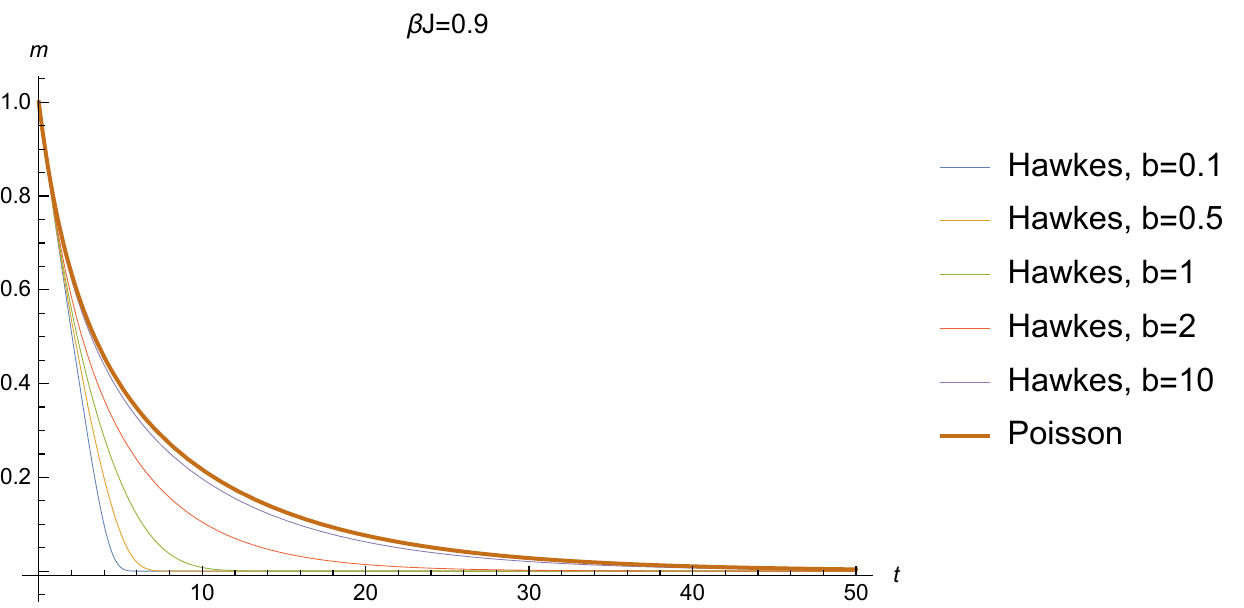}} c) \\
\end{minipage}
\hfill
\begin{minipage}[h]{0.47\linewidth}
\center{\includegraphics[width=1\linewidth]{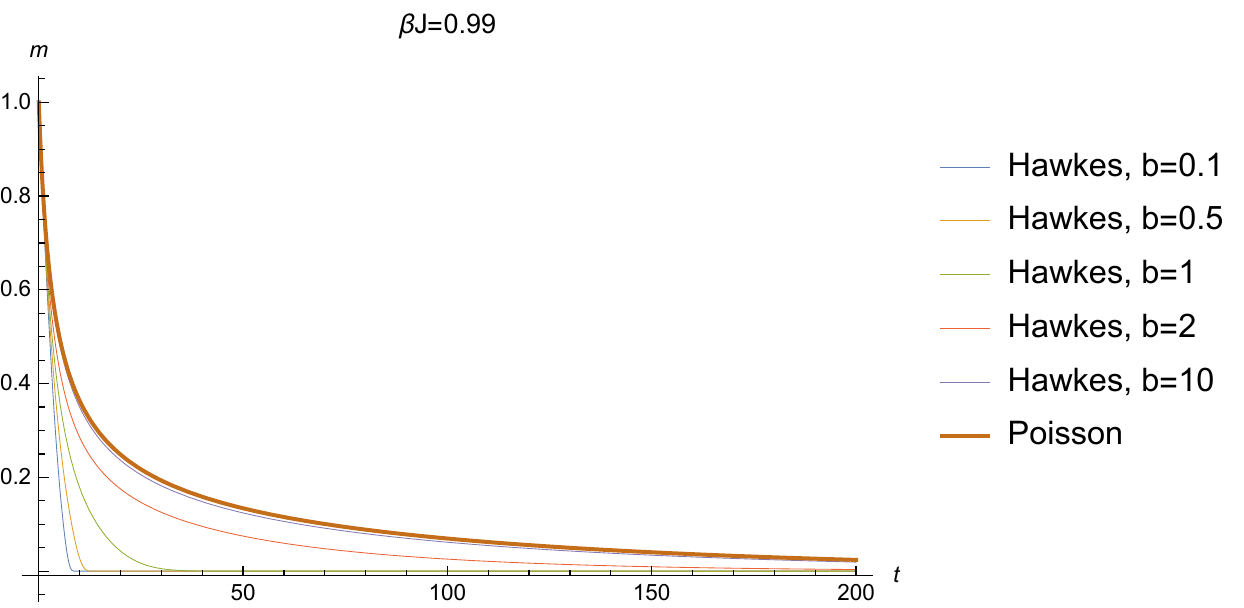}} d) \\
\end{minipage}
\caption{Relaxation from the initial state $m(0)=1$ to the equilibrium $m_{\rm eq}=0$ at $b=0.1, 0.5, 1, 2, 10$ and Poisson case at different temperatures: a) $\beta J=0.1$, b)
$\beta J=0.5$, c) $\beta J=0.9$, d) $\beta J=0.99$. }
\label{para}
\end{figure}

In the high-temperature phase $\beta J <1$ the game has only one equilibrium configuration $m_{\rm eq}=0$. In Fig.~\ref{para} (a - d) we show relaxation from the initial state $m(0)=1$ for a set of values of $b$ at various temperatures $\beta < 1/J$ in comparison to the Poisson case. 
As expected, at all temperatures the self-excited game relaxes faster than in the Poissonian case with relaxation becoming slower with diminishing $b$. In Fig.~\ref{para} (c,d) we also clearly observe the critical slowing down of relaxation characteristic for dynamics in the vicinity of the phase transition point $\beta J = 1$.

In the low-temperature phase $\beta J > 1$ the Curie-Weiss equation has three solutions: the (dynamically) unstable one at $m_{\rm eq}=0$ and a pair of temperature dependent stable equilibria at $m_{\rm eq}=\pm m_0(\beta)$ such that at the critical point $\beta_c=1/J$ one has $m_0(\beta_c)=0$ and at $\beta \to \infty$ one has $m_0 \to 1$.

\begin{figure}[htp!]
\begin{minipage}[h]{0.47\linewidth}
\center{\includegraphics[width=1\linewidth]{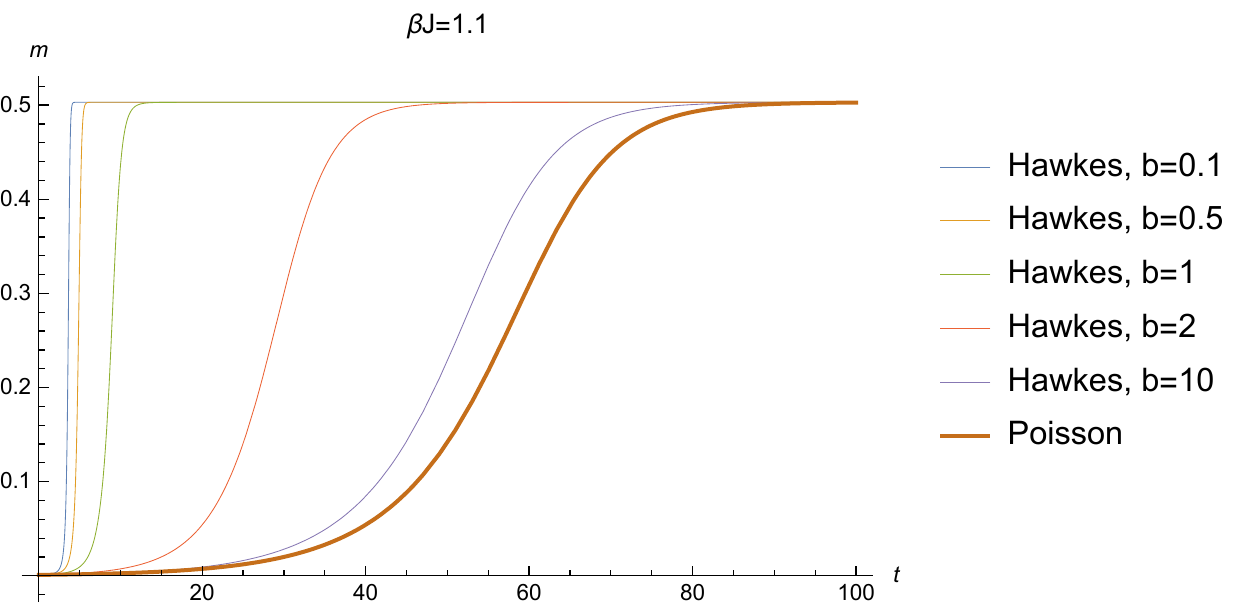}} a) \\
\end{minipage}
\hfill
\begin{minipage}[h]{0.47\linewidth}
\center{\includegraphics[width=1\linewidth]{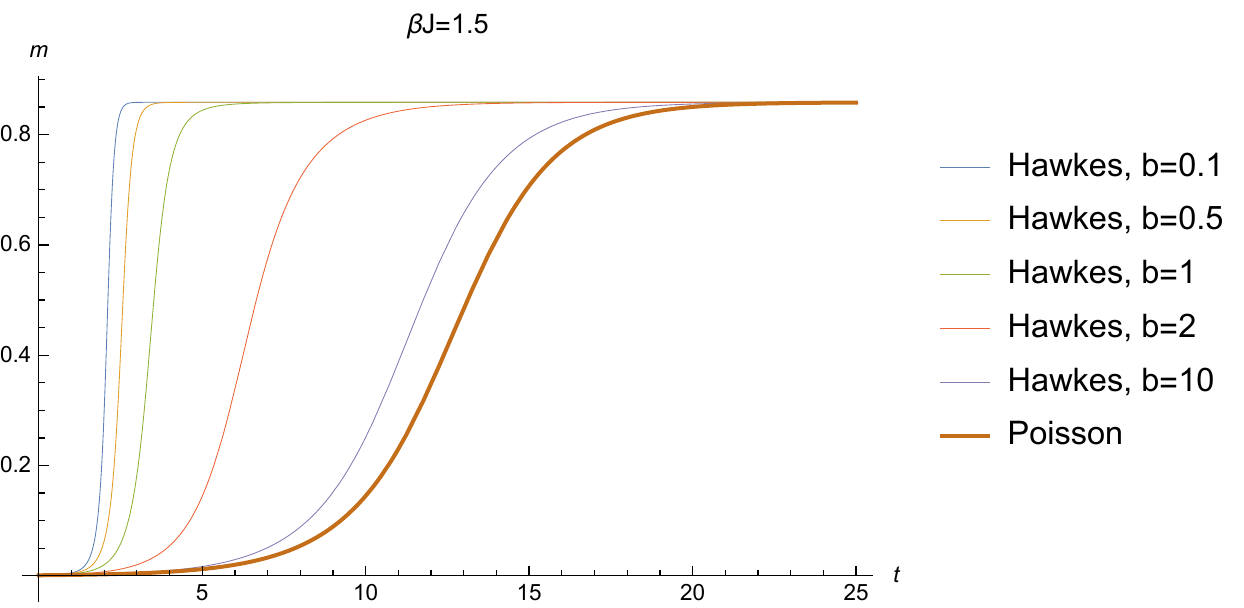}} \\b)
\end{minipage}
\vfill
\begin{minipage}[h]{0.47\linewidth}
\center{\includegraphics[width=1\linewidth]{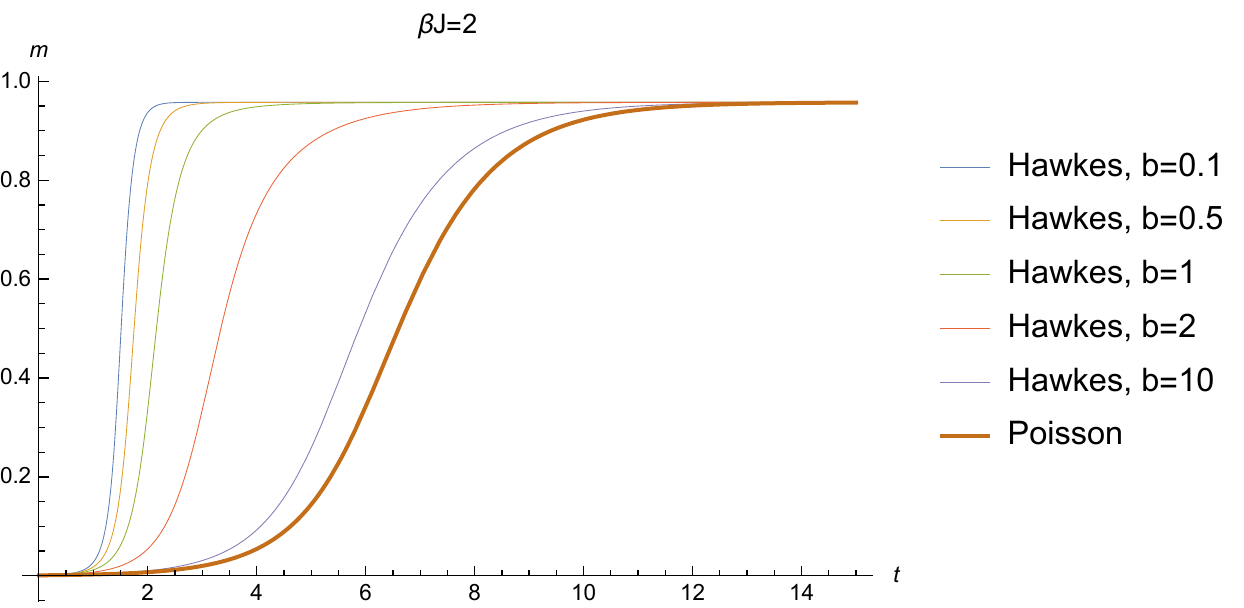}} c) \\
\end{minipage}
\hfill
\begin{minipage}[h]{0.47\linewidth}
\center{\includegraphics[width=1\linewidth]{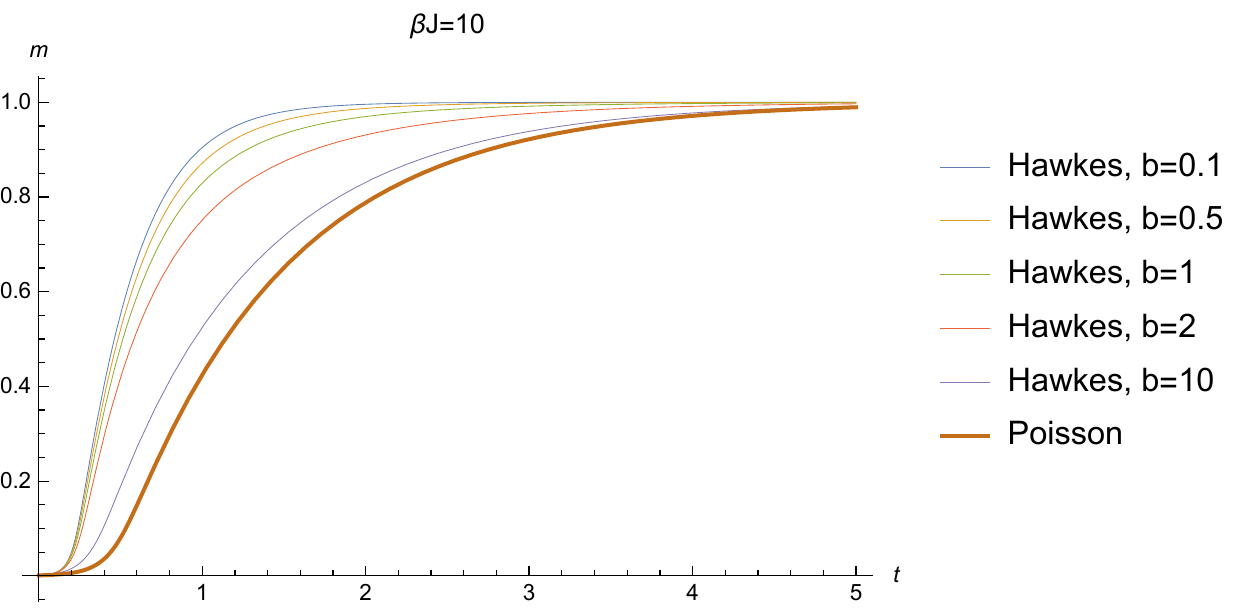}} d) \\
\end{minipage}
\caption{Relaxation from the initial state $m=0$ to the equilibrium $m_{\rm eq} = m_0$ at $b=0.1, 0.5, 1, 2, 10$ and Poisson case at different temperatures: a) $\beta J=1.1$, b)
$\beta J=1.5$, c) $\beta J=2$, d) $\beta J=10$. }
\label{ferro}
\end{figure}

The relaxation to the stable equilibria is illustrated in Fig.~\ref{ferro} (a-d) where for definitiveness we have chosen the equilibrium corresponding to $m_{\rm eq}=m_0$
Similar to the results on above-described case of high temperature equilibrium shown in Fig.~\ref{para} (a-d), the bigger is $b$, the faster is the relaxation to equilibrium.  

\section{Transition between metastable equilibria}\label{trans}

As described in the previous section, in the low - temperature phase $\beta J <1$ the game has two symmetric asymptotically stable equilibria 
$m_{\rm eq}= \pm m_0 (\beta)$. One of the most interesting phenomena in dynamical systems with multiple equilibria is a transition between these attractors such that before settling at $m_{\rm eq}= m_0 (\beta)$ or $m_{\rm eq}= - m_0 (\beta)$ at $t \to \infty$ the systems experiences noise-induced transitions of the type $m_0 (\beta) \leftrightarrow -m_0(\beta)$ at finite times so in this sense at finite times these equilibria are in fact metastable. A brief discussion of a transition of this type in a poissonian Ising game can be found in \cite{Bouchaud2013}. As will be shown below the activity spillover does significantly influence the character of such noise-induced transitions.

To analyse this phenomenon one has to calculate the probability of transition between the states
$(m(t_a),\lambda(t_a))$ and $(m(t_b),\lambda(t_b))$ taking place within the time interval $[t_a,t_b]$. In what follows we shall use a condensed notation $x_{a,b} = (m(t_{a,b}),\lambda(t_{a,b}))$ and fix $[t_a,t_b] = [0,t]$ so that the object of our study is a transition probability
\begin{equation}
\mathcal{P}(x_b, t \vert x_a, 0 ) \equiv P(x_b; t)\Big|_{x(0)=x_a} \label{transprob}
\end{equation}
where $m(0) = \pm m_0 (\beta)$ and $m(t) = \mp m_0 (\beta)$.

\begin{figure}[htp!] \centering{
\includegraphics [scale = 0.9]{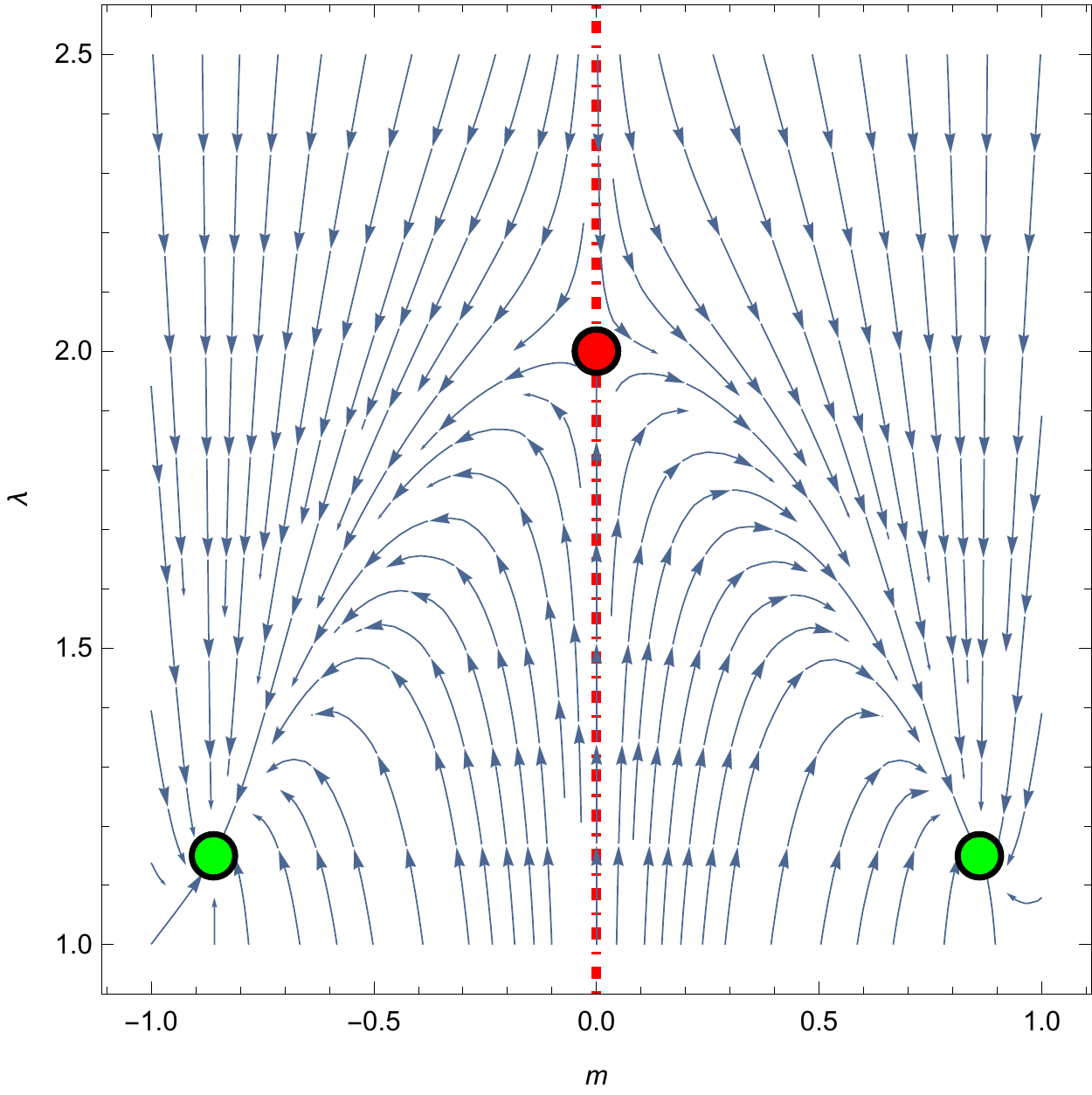}}
\caption{Vector field ${\bf f}=(f_m,f_{\lambda})$ given by equations \eqref{evm} and  \eqref{evel} for $\beta J = 1.5$, $b=2$, $\lambda_0=1$. Two attractors at $m = \pm m_0(\beta)$ (green) are divided by separatrix (dashed red) with a saddle point $m=0$  (red) on it.}
\label{vector}
\end{figure}

The transition probability \eqref{transprob} obeys the following Fokker-Planck equation\footnote{Here and in what follows the indices $i$ and $j$ represent coordinates $m$ and $\lambda$}
\begin{eqnarray}
 \partial_t P & = & \partial_i (f_i P)+\frac{1}{N}\partial_{ij} \left(g_{ij} P\right)\nonumber \\
 f_i & = & \begin{pmatrix}\lambda\left[ m-\tanh(\beta J m) \right] \\  -\lambda \left[ 1-m\tanh(\beta J m) \right]  +b \left[ \lambda-\lambda_0 \right] \end{pmatrix} \nonumber \\
 g_{ij} & = & \begin{pmatrix} \lambda \left[ 1-m\tanh(\beta J m) \right] & -\lambda\left[ m-\tanh(\beta J m) \right] \\  -\lambda\left[ m-\tanh(\beta J m) \right] & \lambda \left[ 1-m\tanh(\beta J m) \right] \end{pmatrix},
\label{FPE}
\end{eqnarray}
that can be derived from the master equation \eqref{eq7}. The Fokker-Planck equation \eqref{FPE} describes Brownian motion in an external vector field $f_i$ in the plane $(m,\lambda)$ subject to noise effects described by $g_{ij}$. Let us stress that in our model, the field $f_i$ is non-gradient since $\partial_{\lambda} f_m \ne \partial_m f_{\lambda}$. Diffusion in such non-gradient fields were conisdered, e.g., in \cite{Maier1993, china}. The vector field $f_i$ is plotted, for a fixed temperature in the low-temperature phase, in Fig.~\ref{vector}, in which we see to attractors corresponding to $m = \pm m_0(\beta)$ and a saddle point at $m=0$.

As the diffusion coefficient in \eqref{FPE} is proportional to $1/N $, in the limit of $N \to \infty$ we can apply to Fokker-Planck equation \eqref{FPE} the standard WKB approximation. Using the notation $ P(x,t) \propto e^{-N W(x,t)} $ we get:

\begin{equation}
\partial_t W(x,t) = -f_i(x(t))\frac{\partial W(x,t)}{\partial x_i} -g_{ij}(x(t))\frac{\partial W(x,t)}{\partial x_i}\frac{\partial W(x,t)}{\partial x_j}.
\end{equation}

The time evolution of the system is then given by following equations\footnote{Below we use the notation $\frac{\partial W}{\partial x_i}=w_i$.}
\begin{eqnarray}
\dot{x_i}(t)+f_i(x(t)) & = & 2 g_{ij}(x(t)) w_j\nonumber \\
\dot{w_i}(t)-w_j\partial_i f_j(x(t)) & = & -w_j \partial_i g_{jk}(x(t))w_k(t).
\label{EOM}
\end{eqnarray}
The system \eqref{EOM} has the following conserved quantity (first integral):
\begin{equation}
f_i(x(t))w_i(t) + g_{ij}(x(t))w_i(t)w_j(t) = E.
\label{curve}
\end{equation}

The leading contribution to the transition probability has the form
\begin{eqnarray}
    \mathcal{P}(x_i,x_f;t) & \propto & e^{-N W}, \nonumber\\
W & = & \int_0^{t}{d\tau} {w_i(\tau)\dot{x_i}(\tau)}-Et = W_0 -Et, 
\label{exp}
\end{eqnarray}

\begin{figure}[htp!]
\begin{minipage}[h]{0.47\linewidth}
\center{\includegraphics[width=1\linewidth]{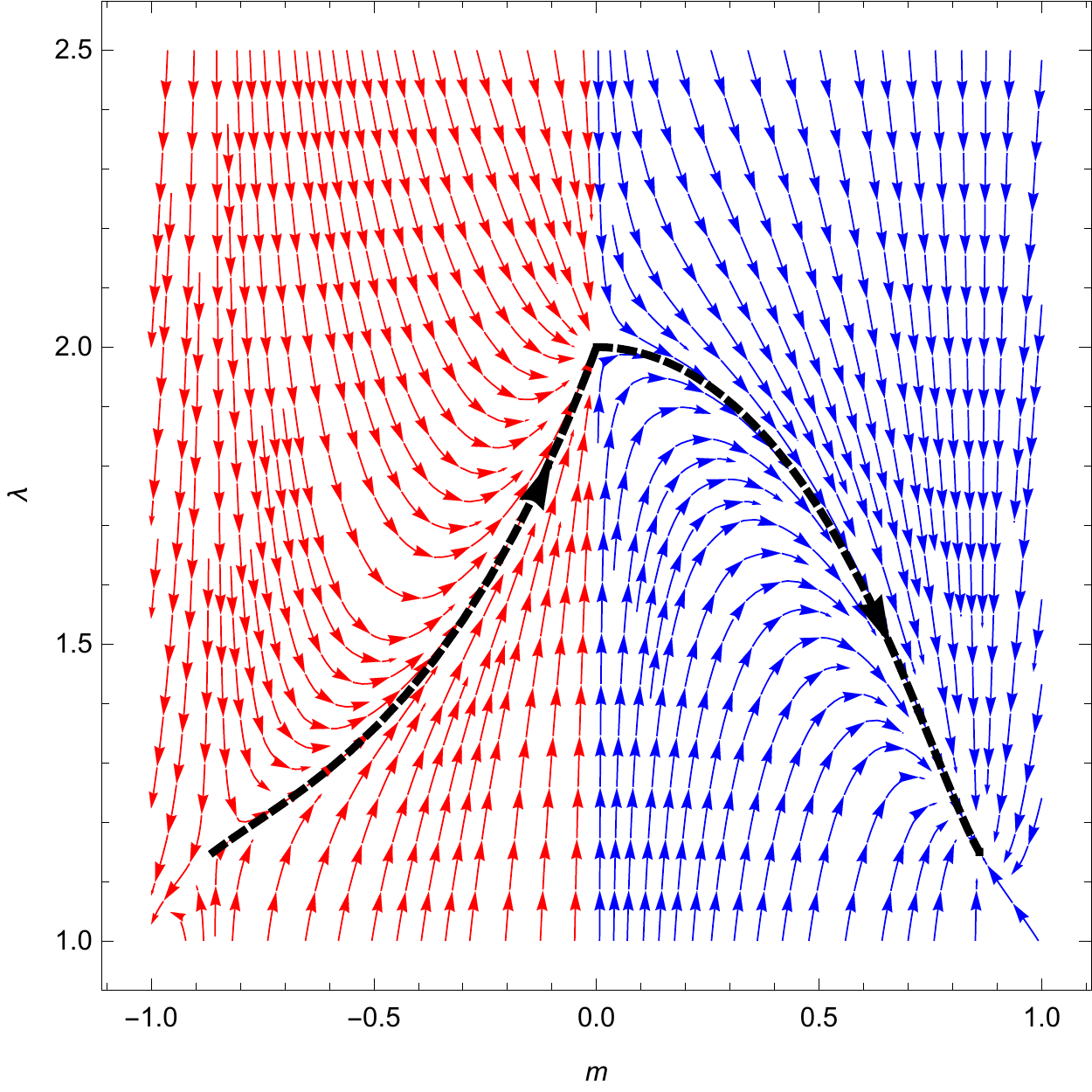}} a) \\
\end{minipage}
\hfill
\begin{minipage}[h]{0.47\linewidth}
\center{\includegraphics[width=1\linewidth]{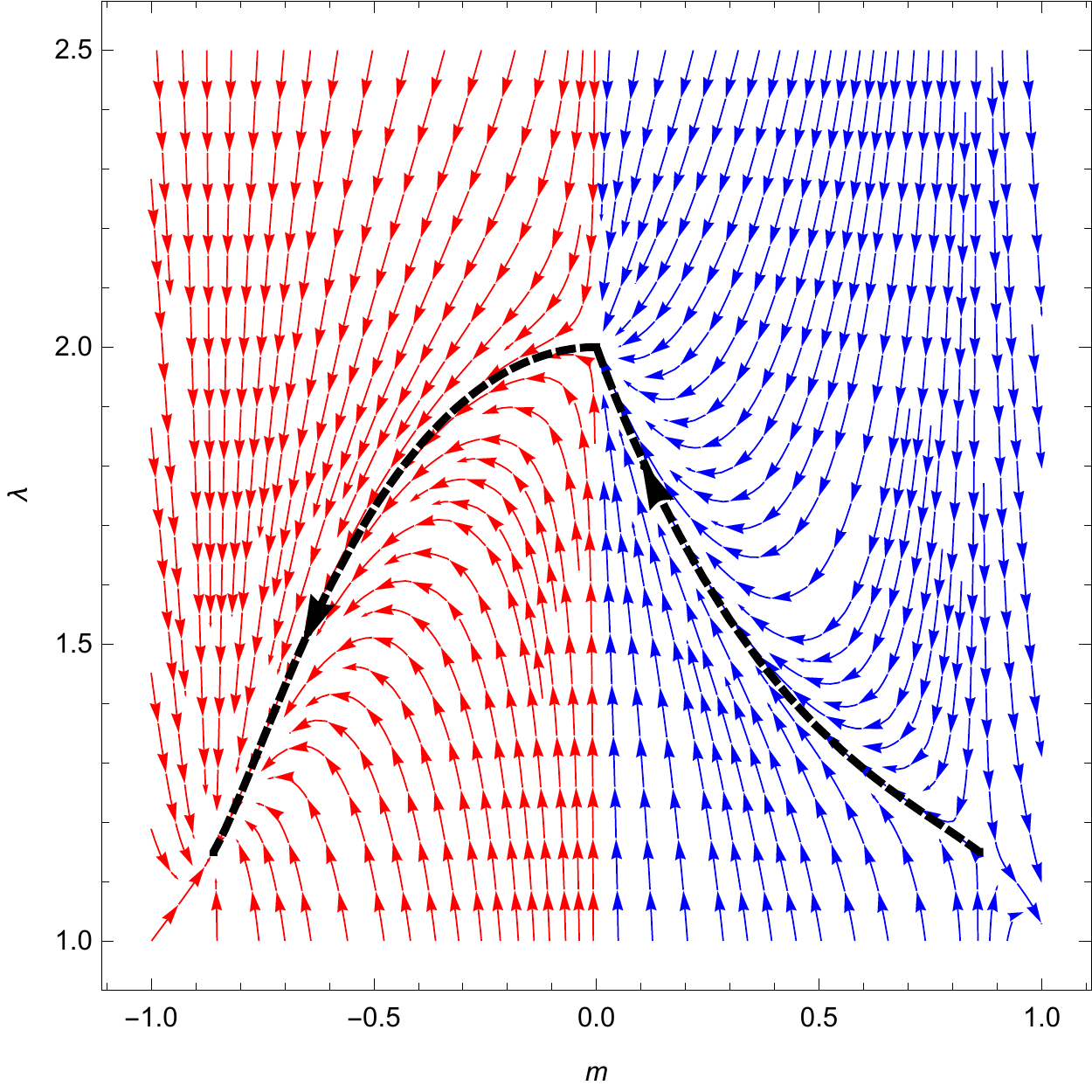}} b)\\
\end{minipage}
\vfill
\begin{minipage}[c]{0.47\linewidth}
\center{\includegraphics[width=1\linewidth]{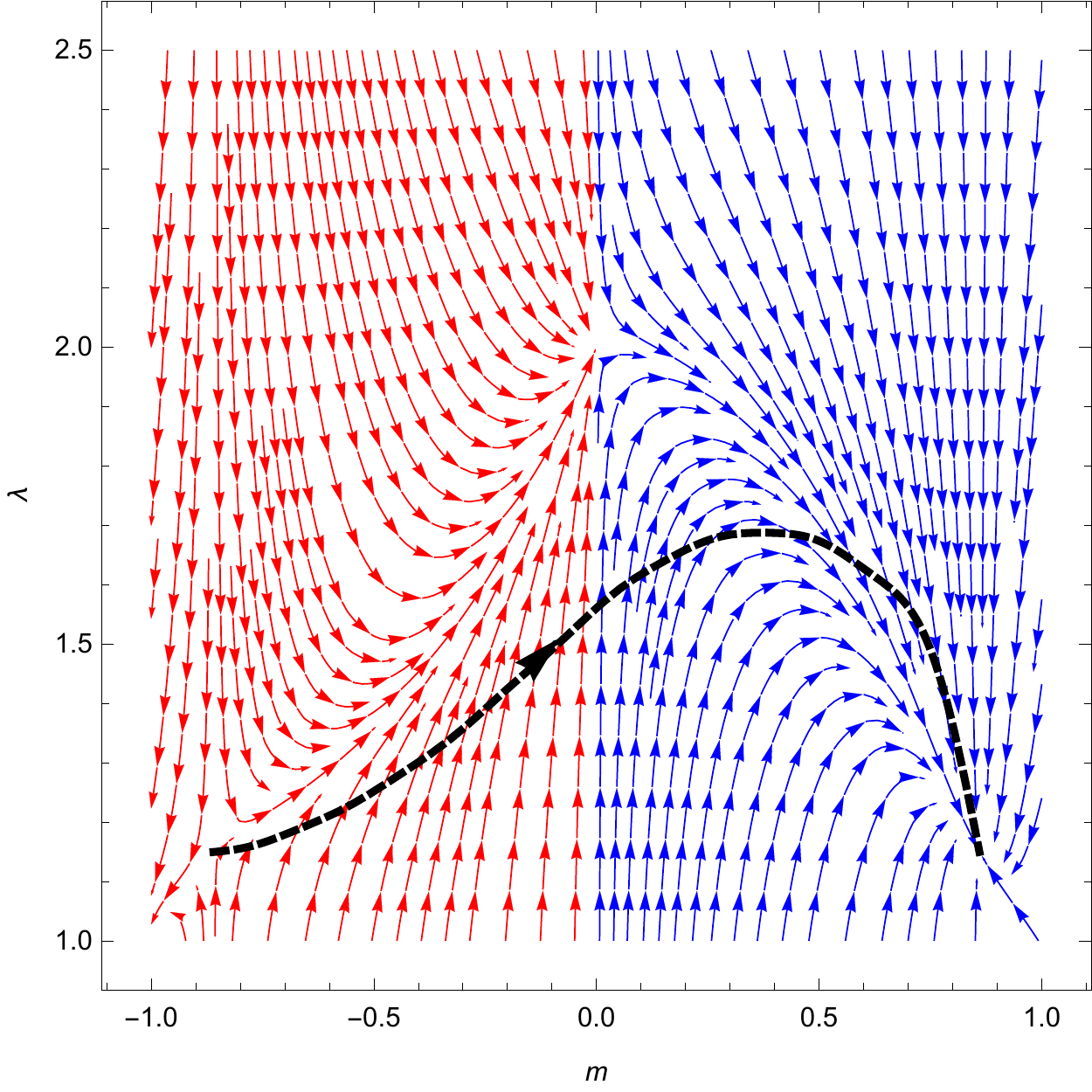}} c) \\
\end{minipage}
\hfill
\begin{minipage}[c]{0.47\linewidth}
\center{\includegraphics[width=1\linewidth]{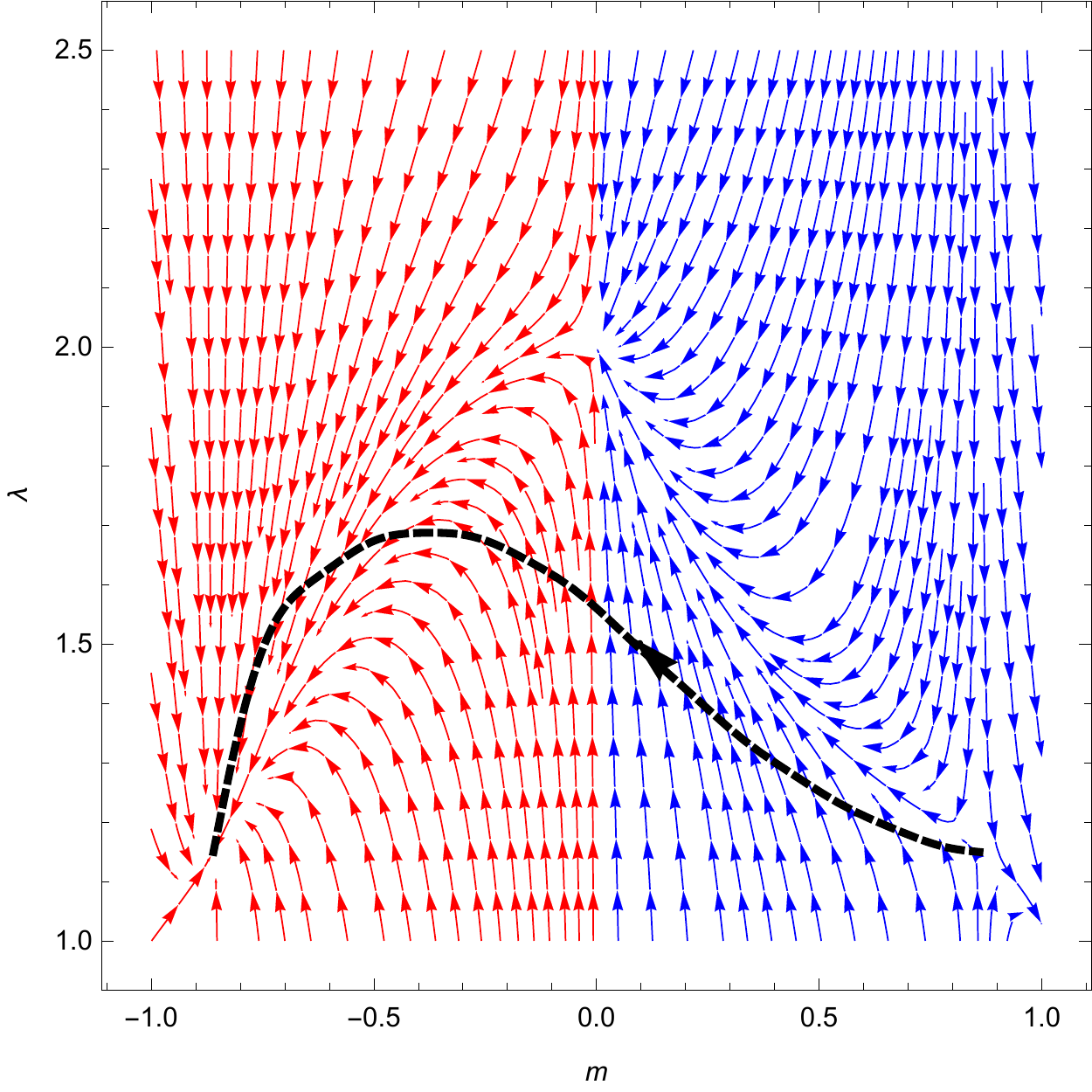}} d) \\
\end{minipage}
\vfill
\begin{minipage}[c]{0.47\linewidth}
\center{\includegraphics[width=1\linewidth]{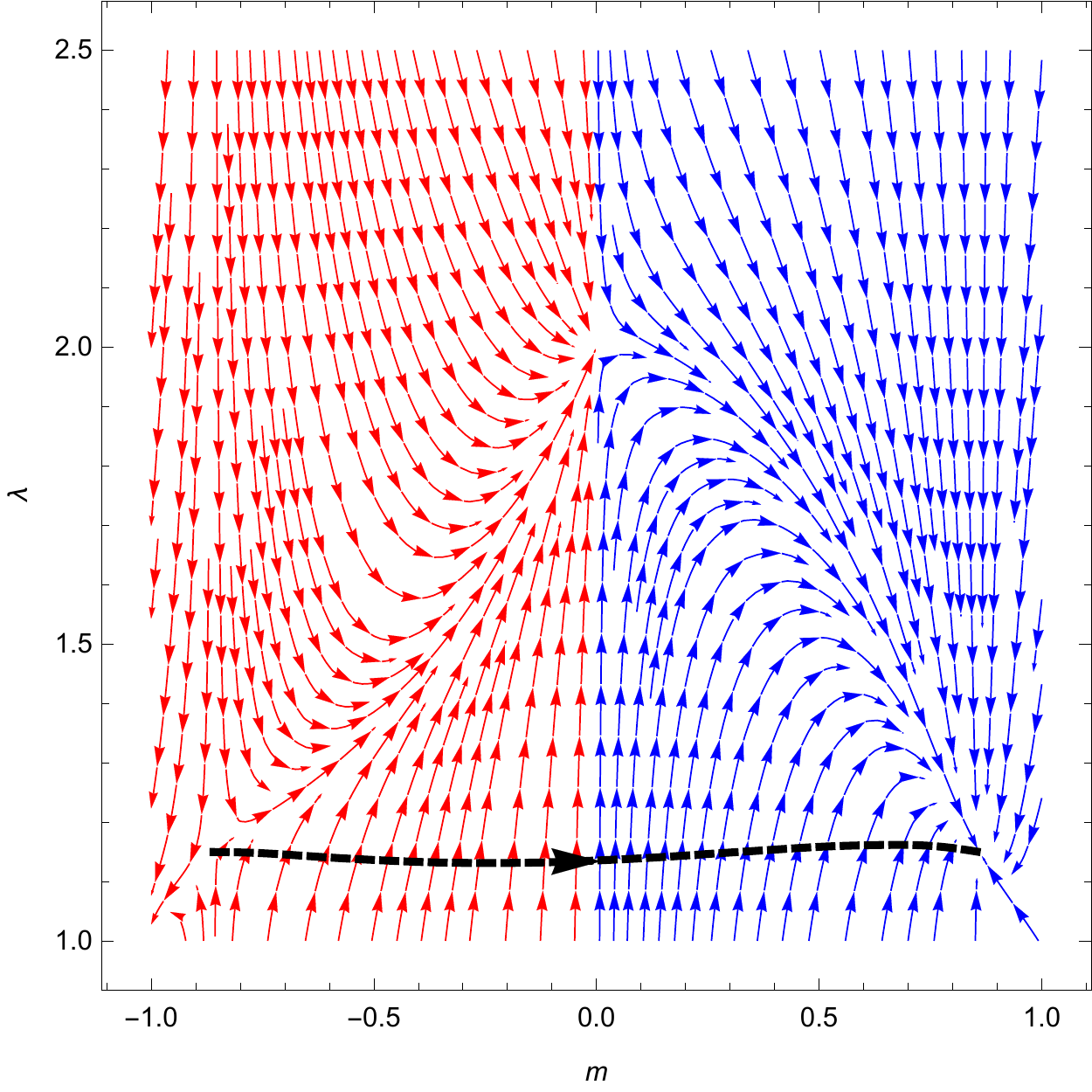}} e) \\
\end{minipage}
\hfill
\begin{minipage}[c]{0.47\linewidth}
\center{\includegraphics[width=1\linewidth]{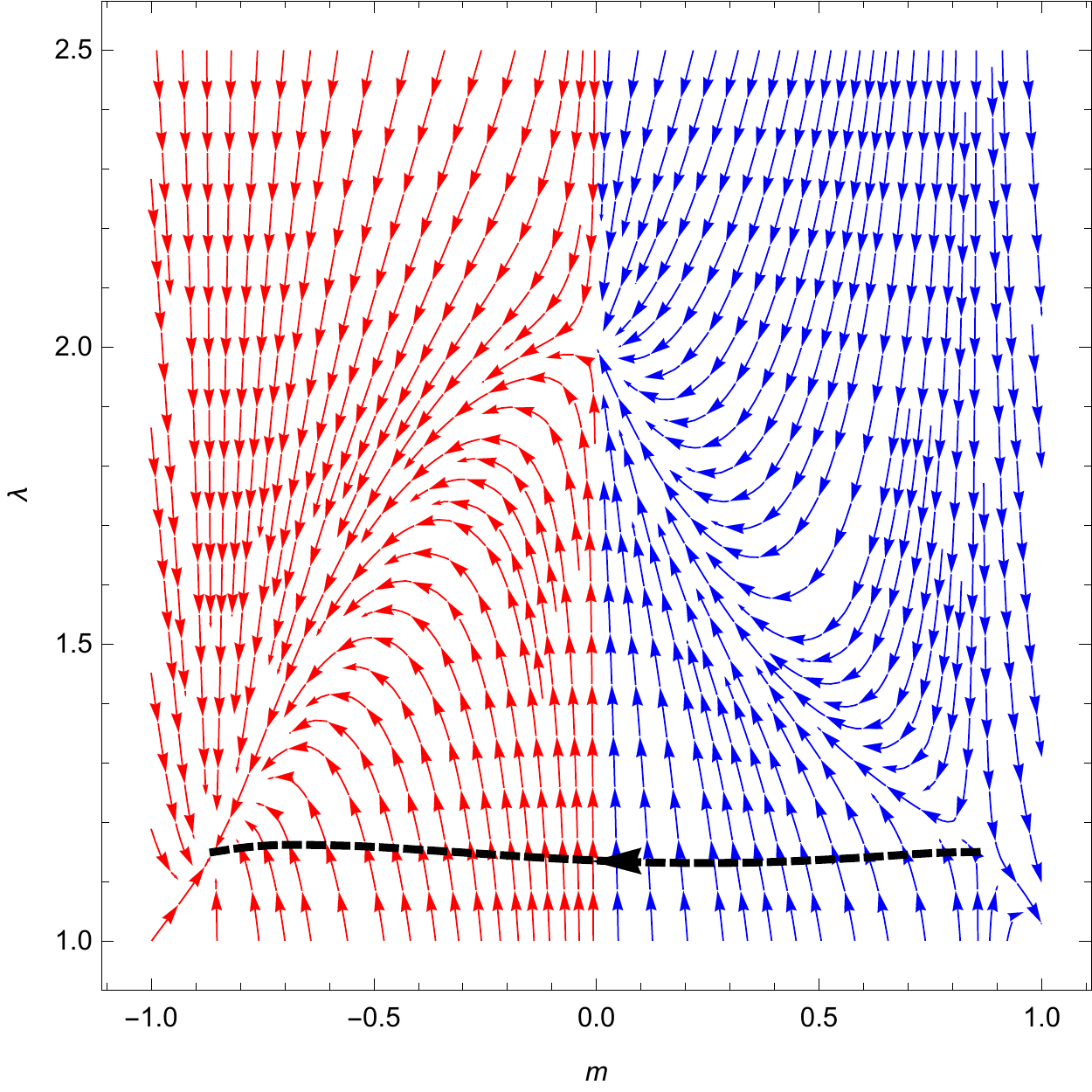}} f) \\
\end{minipage}
\caption{Transition trajectories for $\beta J = 1.5$, $b=2$, $\lambda_0=1$, a, b) $E=0$ ($t=\infty$); c, d) $E=0.01$ ($t \approx 8$); e, f) $E=1$ ($t \approx 0.7$), drawn on top of the vector fields given by corresponding solutions of \eqref{EOM} for $E=0$.}
\label{trajectories}
\end{figure}

The transition trajectory itself is determined by equations \eqref{EOM}, \eqref{curve} and, obviously,  should minimise the trajectory-depended term $W_0$. Examples of trajectories $m_0 (\beta) \to -m_0 (\beta)$  and $-m_0 (\beta) \to m_0 (\beta)$ for various values of $E$ and $t$ are shown in Fig.~\ref{trajectories}. Note the fact that the transition from one attractor to another is accompanied by an increase in the process intensity, so we may expect the transition to be accelerated in comparison with the Poisson process. A noticeable feature of the trajectories shown in Fig.~\ref{trajectories} is their asymmetry with respect to reflection $m \to -m$.

\begin{figure}[htp!] \centering{
\includegraphics [scale = 0.9]{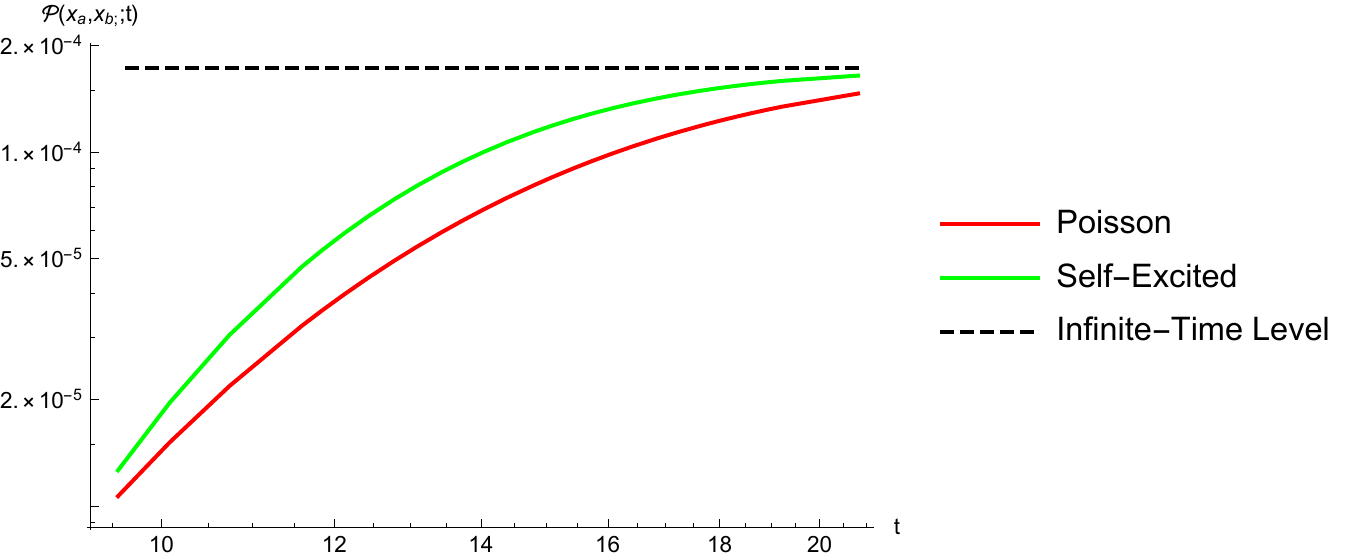}}
\caption{Comparision of transition probability $\mathcal{P} (x_a, x_b; t)$ between self-excited and Poisson cases for $\beta J = 1.5$, $b=2$, $\lambda_0=1$, $N=150$ in log-log-scale.}
\label{compare}
\end{figure}

An important effect of the activity spillover is the amplification of transition probability connecting metastable attractors in comparison to the Poisson process illustrated in Fig.~\ref{compare}. We see that activity spillover leads to exponential amplification of the probability of trajectories 
$\pm m_0 (\beta) \to \mp m_0(\beta)$ at finite times\footnote{The probability itself does however remain small.} 

\section{Conclusion}\label{concl}

The self-excited version of the Ising game taking into account activity spillover on a complete graph was considered. The mechanism of self-excitation was an activity spillover from other agents affecting the decision-taking rate at a node under consideration. The Fokker-Planck equation describing Brownian motion in the $(m, \lambda)$ plane, as well as system of evolution equations for the average choice $m$ and the process intensity $\lambda$ generalising the Curie-Weiss equations, were derived using the master equation formalism. A detailed dependence of the relaxation acceleration on the parameters of the Poissonian memory kernel was studied. We also considered transition from one metastable equilibrium to another in finite time and found out that activity spillover can exponentially accelerate this process.

\begin{center}
{\bf Acknowledgements}
\end{center}

A.L. is grateful to Didier Sornette for a discussion of the problem studied in this paper from which the described research originates.

\newpage

\bibliographystyle{unsrt}
\bibliography{document.bib}

\end{document}